# Reducing the oxygen contamination in conductive (Ti,Zr)N coatings *via* RF-bias assisted reactive sputtering

K. Thorwarth[1]*, M. Watroba[2], O. Pshyk[1], S. Zhuk[1], J. Patidar[1], J. Schwiedrzik[2], J. Sommerhäuser[1], L. Sommerhäuser[1], S. Siol[1]*

[1] Empa, Swiss Federal Laboratories for Materials Science and Technology, Laboratory for Surface Science and Coating Technologies, Dübendorf, Switzerland

[2] Empa, Swiss Federal Laboratories for Materials Science and Technology, Laboratory for Mechanics of Materials and Nanostructures, Thun, Switzerland

E-Mail: kerstin.thorwarth@empa.ch, sebastian.siol@empa.ch

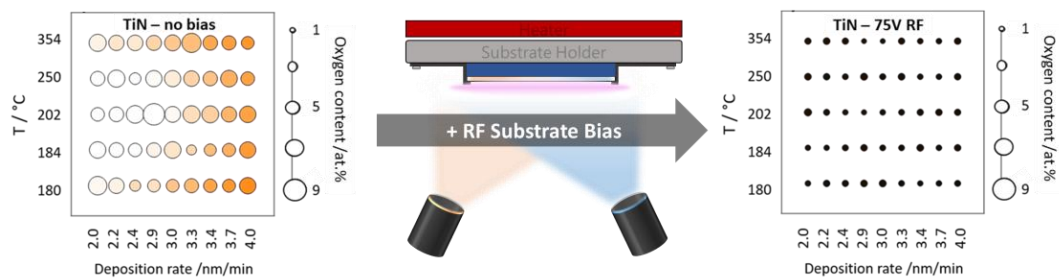

## Abstract
Ternary transition metal nitride coatings are promising for many applications as they can offer improved hardness and oxidation resistance compared to binary counterparts. A common challenge in the deposition of functional nitride thin films is oxygen contamination. Even low amounts of oxygen contamination can adversely affect the functional properties of the thin films, especially hardness and electrical properties. Here, we present a practical approach for the growth of virtually oxygen-free (Ti, Zr)N thin films.

To cover the complete compositional range of (Ti, Zr)N coatings we employ combinatorial reactive co-sputtering. The compositional gradients are complemented by orthogonal deposition temperature gradients to cover broad regions of the synthesis phase diagram. The depositions are carried out with or without applying a low-power radio-frequency (RF) bias voltage to the substrate holder to study the possibility of decelerating energetic oxygen ions and effectively reducing oxygen contamination in the growing film. High-throughput structural analysis and functional property mapping are used to elucidate the synthesis-property relationships in large regions of the synthesis phase diagram. The structural analysis indicates solid solution formation over the entire compositional range, as evidenced by Vegardian lattice scaling, regardless of the applied RF substrate bias. Irrespective of the composition of the films, the application of RF substrate bias leads to a dramatic reduction of oxygen contamination, as demonstrated by X-ray photoelectron spectroscopy (XPS) depth-profile mapping. This is reflected in a significant improvement in the films' conductivity as well as hardness. We demonstrate that the reduction in oxygen contamination is rather intrinsic to the process and not due to changes in the microstructure, which could lead to varying oxidation resistance in ambient conditions. The approach presented here is applicable to both conductive and insulating substrates and provides a practical route to synthesize nitride thin films with improved purity that can be applied in standard sputter chambers and on many different material systems.

## 1. Introduction

Engineering relies on materials as its foundation. With the advent of modern technologies the demand for new or improved functional thin film materials is growing at an unprecedented rate.[1–3] Protective coatings in particular enabled many technological advancements.[4] Especially, nitrides are successfully employed for protective applications due to their high hardness and thermal stability.[5] Among those, binary systems like TiN and ZrN, and ternary systems like TiAlN are the most industrially established examples. TiN and ZrN coatings are widely known as standard coatings for cutting tools and have been studied for more than 30 years.[6–9] TiN is nowadays used for in many applications ranging from hard protective coatings as well as for medical and semiconductor applications. However, TiN has one serious disadvantage. It cannot be used in air at temperatures above 723 K, e.g. aerospace and high-temperature corrosion applications, because it easily oxidizes to titanium dioxide ($TiO_2$) resulting in significant degradation of its performance.[10] To overcome this drawback, several attempts have been made to improve the oxidation resistance of TiN by alloying it with other elements. Examples of this modification are (Ti, Al)N, (Ti, Al)(O, N), (Ti, Si)(C, N), (Ti, Hf)N, or (Ti, Al, Zr, Si)N[11–14], which all demonstrated an improved high temperature performance in comparison to TiN. Upon alloying TiN with Zr, Zr can completely substitute Ti in the cation sublattice of the NaCl-type (Space group 225) lattice, while maintaining the cubic close-packing structure. This leads to a noticeable improvement of the diffusion barrier performance and thus corrosion resistance of TiZrN compared to its binary counterparts.[15] In addition, solid-solution strengthening upon alloying also enhances the hardness of the film by restricting the movement of dislocations.[16]

TiN as well as ZrN show unusual electron transport as well as optical and mechanical properties compared to their metal counterparts.[17] Both can host different amounts of nitrogen, with respect to the stoichiometric compounds. A variation of nitrogen content does not affect the crystal structure (only a small change in lattice parameters)[18], but significantly changes the functional properties such as hardness and resistivity[19]

The effect of nitrogen vacancies was studied in detail for several nitrides such as ZrN[20] and TiN[21], and was found to not only result in poorer functional properties, but also lower crystalline quality. Similarly, oxygen contamination plays a critical role as it often adversely affects the functional properties of the thin films. This can be attributed to the formation of oxygen-related defects, such as oxygen interstitials or $TiO_x$ and $ZrO_x$ secondary phases, which can act as scattering centers for electrons and lead to an increase in resistivity. Additionally, the NaCl lattice expands when nitrogen is replaced by an oxygen atom in the anion sub-lattice, leading to the formation of Ti(O, N) or Zr(O, N) oxynitrides, respectively. The latter also causes a transition from metallic to semiconducting behavior.[22]

Consequentially, oxygen contamination is a major challenge in the deposition of functional nitrides. This is especially important for production environments where ultra-high vacuum systems and high-purity precursors are directly linked to significantly higher investments. Oxygen contamination in sputtered nitride films often results from residual oxygen in the sputter chamber as well as impurities in the sputtering gas. Although such oxygen contamination in the sputter gas is not detectable by standard residual gas analysis, the high sticking coefficient of $O_2$ results in a high oxygen contamination level even at very low impurities in the gas. This is especially the case for films grown at low deposition rates. A positive effect of a radio frequency (RF) bias on the oxygen contamination level in sputtered InN films was previously reported by Westra et al.[23] Already in 1965, the influence of a negative bias to the substrate was found to remove residual oxygen contamination in sputtered metallic Ta, Nb, Al or Mo films.[24] In this earlier work, it was speculated that the sufficient bias allows for the removal of adsorbed impurities originated at the cathode and traveled to the substrate during sputtering, since the ionic bombardment of a sputtered film results in a preferential removal of adsorbed gas. These promising results motivate a systematic investigation if low power RF substrate bias potentials can be used to reduce the residual oxygen incorporation during sputter deposition of functional nitride thin films.

In this work we demonstrate that by applying an RF substrate bias potential during sputtering, the oxygen contamination in (Ti, Zr)N films on insulating or electrically floating substrates can be reduced to values close to zero (i.e. below the detection limit of our X-ray photoelectron spectroscopy (XPS) measurements). We conduct a full combinatorial phase screening of (Ti, Zr)N covering large regions of the synthesis phase space. By employing automated characterization of the structure, composition, and functional properties, we show that depending on the deposition conditions, the functional properties can be significantly improved over conventional deposition with a floating substrate potential. Specifically, the hardness and conductivity of TiN, ZrN, and (Ti, Zr)N films deposited using the RF substrate bias potentials are markedly higher. The results of this work are easily transferrable to many other material systems and can be employed using standard deposition equipment.

## 2.     Experimental

Thin films of (Ti, Zr)N were deposited by reactive magnetron co-sputtering from two magnetrons in confocal sputter-up geometry in an Ultra-High Vacuum (UHV) chamber (AJA ATC 1500 F) using an unbalanced closed field magnetron configuration to minimize plasma heating.[25,26] 2'' metallic targets were used for the deposition: Ti (99.995% purity, HWM Hauner) and Zr (99.995% purity, HWM Hauner). A more detailed description of the magnetron sputtering setup is given by Trant et al.[26] High-quality borosilicate glass (Corning, Eagle XG) is used as a substrate because it allows for the simultaneous determination of optical and electrical properties of deposited films in combinatorial studies. Glass substrates (1.1 mm thickness) of 50.8 mm × 50.8 mm size were ultrasonically cleaned in acetone–ethanol mixture for 5 min. After drying using $N_2$ the glass was partially glued to a custom-made substrate holder using silver paste. As only one side of the glass was in direct contact with the heater, a temperature gradient was achieved from ~181°C (cold side) to ~354°C (hot side). The temperature calibration was performed using type k thermocouples, glued to the glass substrate as well as infrared imaging through a ZnS viewport. The base pressure before sputtering was below $10^{-5}$ Pa. The process pressure was kept at 0.5 Pa with an Ar flow of 18 sccm (to the chamber) and 8 sccm $N_2$ flow was introduced directly to the magnetrons resulting in a fully poisoned target. The deposition was carried out without substrate rotation at a sputter angle of approximately 30°. This results in combinatorial sputter flux gradients, which allows us to co-vary temperature and metal-to-nitrogen ratio in the binary TiN and ZrN systems. For the (Ti,Zr)N alloys it allows co-variation of temperature and composition. For the binary samples (TiN and ZrN), the power density was kept constant at 9.88 W/cm$^2$. For the ternaries, the power on both targets was adjusted to keep the deposition rate constant across the substrate for each sample on the combinatorial library (6 nm/min).

The insulating substrate was mounted on a substrate holder, which was either in the floating state which corresponds to a floating potential of approximately -18 V or with an applied moderate RF bias with a frequency of 13.56 MHz and a potential of -75 V. Lower bias voltages resulted in an unstable plasma while a higher bias voltage was not favorable, because a higher bias voltage would also result in ion implantation resulting in a significant change in microstructure and excess compressive stress.[27] To ensure no active re-sputtering of the film, a TiN-coated glass slide was masked and exposed to the -75 V RF substrate bias voltage for 120 min in the presence of the Ar/N process gas. No etching of the TiN was observed under such conditions.

After deposition, films were cooled down in a Nitrogen atmosphere. All deposition parameters are also listed in Table 1. The results were reproduced in a second deposition tool (AJA Orion 8) with comparable geometry and identical deposition parameters. For this study no gas purification was employed in either of the deposition systems.

**Table 1.** Deposition parameters for TiZrN thin films.

| Parameter | TiN | ZrN | TiZrN |
|---|---|---|---|
| **Target-substrate distance (cm)** | | 11.5 | |
| **Sputter angle (°)** | | 25 | |
| **Target power density (Wcm$^{-2}$)** | 9.88 | ~9.88 | Ti: 6.6-13.6; Zr: 2.5-5.5 |
| **Deposition rate (nm/min)** | 2-4 | 3.4-7 | 6 |
| **Working pressure (Pa)** | | 0.5 | |
| **Ar/N$_2$ flow rate (sccm)** | | 18/8 | |
| **Substrate temperature (°C)** | | 181-354°C | |
| **Substrate bias** | | floating / -75 V RF | |

The cation concentration and the film thickness of the libraries were quantified using fully automated X-ray fluorescence (XRF) mapping (Fischerscope XDV SDD, equipped with an Rh X-ray source). The film thickness measurements were confirmed with a profilometer (Dektak XT, Bruker). Additionally, selected samples were measured by X-ray photoelectron spectroscopy (XPS) (PHI-Quantera, ULVAC-PHI Inc.) with Al Kα radiation source (hν = 1486.7 eV). The composition of the films, including the oxygen contamination, was measured by performing sputter-etching by means of a 1 keV Ar beam scanned over an area of 2 x 2 mm$^2$ to acquire sputter-depth profiles. Charge neutralization was performed using a dual-beam neutralizer for the films on glass substrates. Additional selected film libraries were transferred to the XPS chamber from the deposition chamber using a custom-built UHV-transfer cart. The pressure was maintained below 10$^{-4}$ Pa throughout the transfer process, avoiding surface contamination from ambient air.[25] Semi-quantitative composition analysis was performed using instrument-specific relative-sensitivity factors after Shirley background subtraction.

The structure and phase composition of the films were characterized using X-ray diffraction (XRD) mapping (D8 Discover, Bruker) with Cu Kα X-ray source (λ = 1.54 Å) and Bragg Brentano geometry. The interplanar spacing $d$ is determined using Bragg's law $\lambda=2d\sin\theta$ where $\lambda$ is the X-ray wavelength, $d$ is the spacing of the diffracting planes, and $\theta$ is the angle between the incident rays and the diffracting planes, also known as the Bragg angle.

The resistivity was determined using a custom-built, fully automated Four-Point Probe measurement system with a SP4 probe head (Microworld) and a 2612B source meter (Keithley). The probe head has four tungsten carbide pins with a 125 µm tip radius, 45 gram spring pressure per pin, and 1 mm spacing between pins. Current is passed through the outer pins while the voltage is measured between the two inner tips. The voltage drop across the sample was measured at different source currents ranging from -10 µA to 10 µA in 2.5 µA steps. The sheet resistance was calculated according to

$$R_S = \frac{\pi}{\ln(2)} * \frac{V}{I} = \frac{\pi}{\ln(2)} * \frac{N \sum(I_i V_i) - \sum I_i \sum V_i}{N \sum I_i^2 - (\sum I_i)^2} \quad \text{(Eq. 1)}$$

where $N$ denotes the number of steps and $\frac{V}{I}$ was approximated by the slope of the least squares linear regression of the measurement values. Resistivity ($\rho$) was then calculated by multiplying the Sheet Resistance $R_S$ with the film thickness d according to: $\rho = R_S * d$.

Nanoindentation was performed on selected samples using an *ex-situ* scanning nanoindenter setup built in-house (Alemnis AG) equipped with a Berkovich diamond tip (Synton MDP). In this experiment, a 3 x 3 matrix of indents with a spacing of 20 µm was performed in displacement-control mode with the displacement rate of 5 nm/s on each of the 45 positions in the library. A maximum indentation depth of less than 10% of the coating thickness was ensured to avoid substrate influence on the measured mechanical properties. The calibration of tip area function was performed before testing each materials library based on continuous stiffness measurement (CSM) tests on a reference fused silica sample. The load-displacement data was evaluated using AMMDA software (Alemnis). The hardness was calculated from the unloading curve using the well-established Oliver–Pharr method.[28]

## 3. Results and discussion

### 3.1 Compositional and structural analysis

#### Oxygen contamination studies

Oxygen contamination in TiN has been widely studied in the past. Pristine TiN typically appears bright golden in color whereas even miniature oxygen contamination can result in an apparent color change to a more bronze/brownish color.[29–31] This feature, together with the high oxygen affinity make TiN a perfect candidate for this study. Strikingly, the TiN films grown for this study showed a very apparent change in color for different deposition conditions. TiN films grown without RF bias appeared brownish, while those with RF-bias had a golden color. This indicates a reduction in oxygen contamination when an RF substrate bias is employed. To prove the latter, the effect of the applied RF substrate bias on the composition of the TiN films was studied by performing XPS analysis.

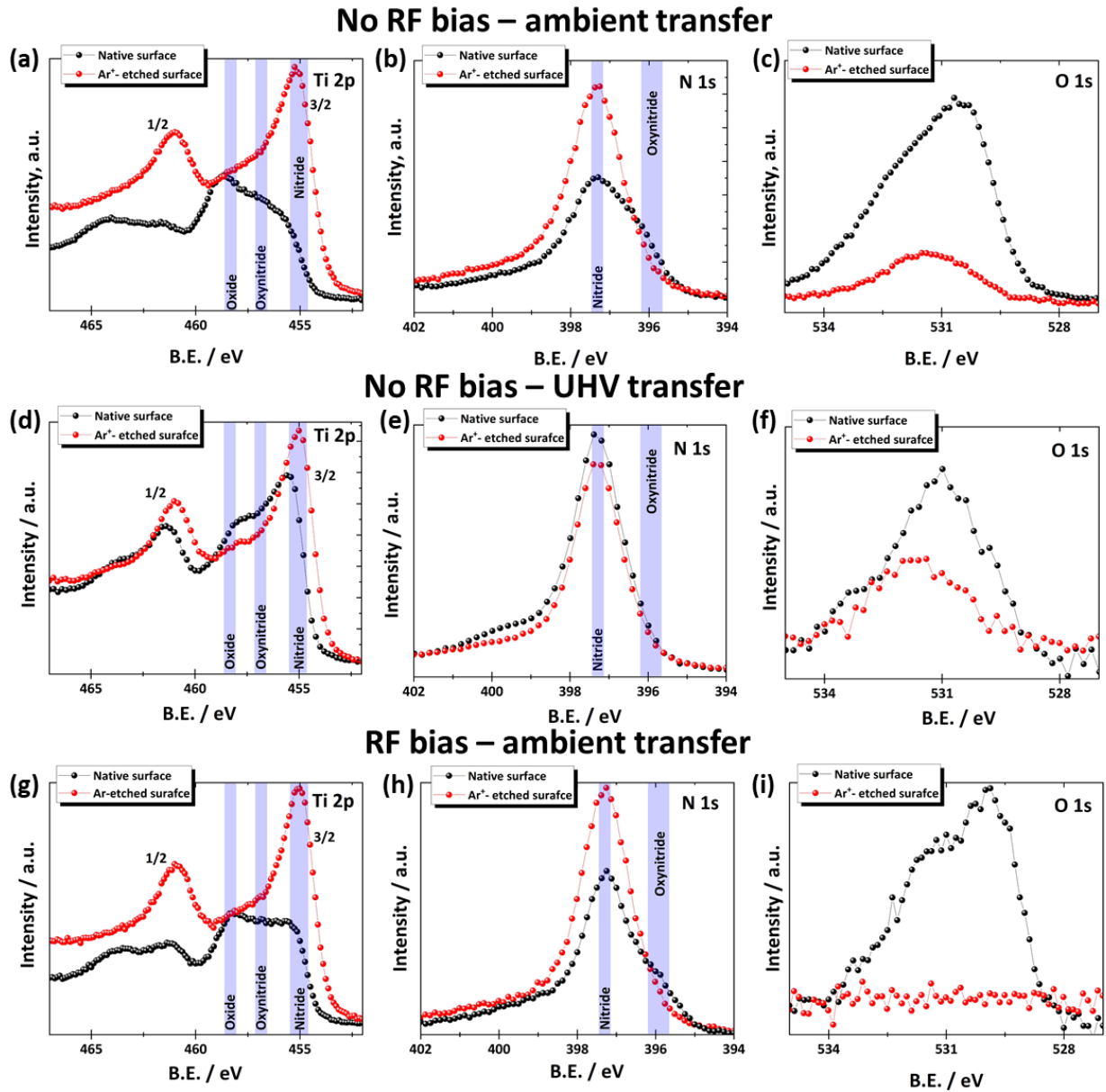

**Figure 1:** XPS core-level Ti 2p, N 1s, and O 1s spectra acquired before (black lines and symbols) and after (red lines and symbols) Ar⁺ sputter-etching of TiN films deposited without (a-f) and with applying an RF bias (g-i). XPS measurements for TiN films grown without RF substrate bias (d-f) were performed after the UHV transfer of the sample from the deposition chamber to the XPS chamber.

**Figure 1** shows the most relevant core levels, i.e. Ti 2p, N 1s, and O 1s for TiN films grown with and without applied RF substrate bias. The Ti $2p_{3/2}$ peak from a TiN film grown without RF substrate bias (Fig. 2a) contains three components at binding energies matching to Ti atoms participating in Ti-N, Ti-O-N and Ti-O bonding in nitride, oxynitride, and oxide phases, respectively. Ar⁺ sputter-etching results in a slight increase in Ti-N component intensity. The intensity of the corresponding Ti-N component from N 1s spectra increases after Ar⁺ etching (Fig. 2b). The analysis of the spectra acquired from the UHV-transferred TiN film grown without RF substrate bias shows that Ti-N components in Ti 2p and N 1s spectra have the highest intensity among all components (Fig. 2d-e) before and after Ar$^+$ sputter-etching. Simultaneously, the O 1s peak intensity decreases after Ar$^+$ sputter-etching (Fig. 2f). In contrast, the Ti 2p$_{3/2}$ peak acquired from the air-exposed TiN film grown with RF substrate bias (Fig. 2g) demonstrates three components, i.e. Ti-N, Ti-O-N and Ti-O components, with relatively equal intensity. However, the Ti-N component from the Ti 2p$_{3/2}$ peak becomes dominant upon the removal of the surface oxide layer and other contaminations by means of Ar$^+$ sputter-etching. The intensity of the oxynitride peak in N 1s spectra (Fig. 2h) decreases upon Ar$^+$ sputter-cleaning while the nitride component increases. Importantly, oxygen was not detected at all after Ar$^+$ sputter-etching of the film grown with RF substrate bias (Fig. 2i) even though the sample was transferred to the XPS chamber in the ambient atmosphere. XPS core-level Zr 3d, N 1s, and O 1s spectra for ZrN films are provided in the Supporting Information (**Fig. S1**), showing a similar trend except for a low intense oxygen signal detected from the sample grown with RF substrate bias.

The surface composition of TiN and ZrN thin films was further analyzed for a semi-quantitative assessment of the oxygen contamination in the films bulk. The results are given in **Table 2**. XPS composition analysis reveals a significantly higher level of oxygen in TiN samples grown without RF substrate bias (up to 8 at.%) compared to samples grown with a -75 V RF substrate bias (below XPS detection limit) after Ar$^+$ sputter-etching. To confirm that the presence of oxygen in the samples grown without RF substrate bias originated from bulk contamination rather than ambient oxidation, a UHV transfer of the TiN layer directly after growth from the vacuum chamber into the XPS, while maintaining a vacuum, was performed. Without UHV transfer, the oxygen content was measured at 8.08 at.% after sputter profiling. With the UHV transfer, the sample prepared under the same process parameters exhibited an oxygen content of 3.52 at.% after 2 minutes of Ar$^+$ sputter-etching using 1 kV Argon. Thus it can be concluded that oxygen was incorporated during film growth and not by post-oxidation along the grain boundaries after being exposed to air. The latter scenario is usually the case for transition metal nitride thin films with columnar microstructure containing inter- and intra-columnar voids.[32,33] Therefore, a morphology effect can be ruled out. TiN sample prepared with applied RF substrate bias showed an oxygen content below the detection limit of XPS. XPS analysis of ZrN thin films shows a similar trend (Table 2), but the sample grown with RF bias contains a small amount of oxygen due to its slightly under-stoichiometric composition, which is more prone to oxidation.

**Table 2.** XPS-derived composition of TiN and ZrN films grown without and with RF substrate bias, where Me stands for Ti or Zr.

| Sample | | Me, at % | N, at % | O, at % | C, at % |
|---|---|---|---|---|---|
| **TiN- no bias, ambient-transfer** | Native surface | 23.6 | 22.4 | 27.1 | 26.9 |
| | Ar$^+$ sputter-etched | 43.5 | 39.2 | 8.1 | 9.2 |
| **TiN- no bias, UHV transfer** | Native surface | 39.1 | 52.0 | 6.7 | 2.3 |
| | Ar$^+$ sputter-etched | 47.1 | 49.4 | 3.5 | 0.0 |
| **TiN-RF bias, ambient-transfer** | Native surface | 27.1 | 29.9 | 18.8 | 24.2 |
| | Ar$^+$ sputter-etched | 48.7 | 50.1 | **0.0** | 1.2 |
| **ZrN- no bias, ambient-transfer** | Native surface | 28.0 | 15.3 | 34.8 | 21.8 |
| | Ar$^+$ sputter-etched | 54.9 | 38.7 | 6.5 | 0.0 |
| **ZrN- RF bias, ambient-transfer** | Native surface | 18.3 | 7.3 | 27.9 | 46.5 |
| | Ar$^+$ sputter-etched | 57.6 | 40.1 | **2.3** | 0.0 |

Additional measurements were performed on TiN films grown with similar deposition conditions on a different sputter chamber (AJA Orion 8) to verify and reproduce the effect of the RF substrate bias, but also to exclude chamber-specific influences like e.g. contaminated gas lines or smaller vacuum leaks. The films deposited in this chamber show a very similar behavior. Here, the TiN film grown without RF substrate bias contains 13.7 at.% of oxygen after Ar$^+$ sputter-etching, whereas, the reference TiN deposited with RF substrate bias contains only 0.4 at.% of oxygen. The detailed results from this XPS analysis are given in the supporting information (**Fig. S2-S3**).

Compositional and structural analysis of (Ti,Zr)N libraries

To extend the investigation to a broader compositional range, Ti$_x$Zr$_{1-x}$N thin films with a composition of $x$ ranging from 0 to 1 were synthesized by reactive co-sputtering with and without applying an RF bias to the substrate holder. The combinatorial libraries were synthesized to exhibit composition gradients on the x-axis and temperature gradients on the y-axis. Sputter depth profile mapping on selected libraries showed only small variations of the cation/anion ratio. ZrN requires a slightly higher nitrogen partial pressure during the deposition. In this study we optimized the conditions for TiN resulting in slightly understoichiometric films for high Zr concentrations. The Ti/Zr cation ratio $x$, as well as the film thickness, was determined by XRF as a reference for subsequent characterization. The surface and bulk composition of the film is nearly identical indicating no preferential oxidation of the films in air (**Figure S4**).

To illustrate the phase evolution in $Ti_xZr_{1-x}N$ thin films, combined XRD and XRF measurements were taken for samples synthesized without RF and with RF substrate bias and compared to the respective simulated reference patterns (see **Figure 2**). In **Figure 2a** individual XRD patterns for TiN, ZrN, and $Ti_{0.5}Zr_{0.5}N$ grown with and without RF substrate bias are presented along with reference patterns obtained from the Materials Project repository[34]. All films exhibit preferential out-of-plane orientation. The TiN film shows a mixture of (002) and (111) out-of-plane texture while the ZrN film grows predominantly with (002) texture (Fig. 5a). For TiN it is well established that the texture develops through a complex interplay between collision-dependent Ti adatom mobility, favoring either a (002) or a (111) texture depending on the substrate temperature, ion-to-neutral flux ratio and ion-irradiation-induced lattice distortion. It was found that higher ion energies favor a (002) texture[35]. This effect can be observed for both TiN and ZrN. The films deposited with RF bias on the substrate holder grow with a higher preference for the [002] direction. The crystal size (calculated using the Scherrer equation) in [002] direction increases for all samples with increasing temperature owing to increased mobility (**Figure S5**).

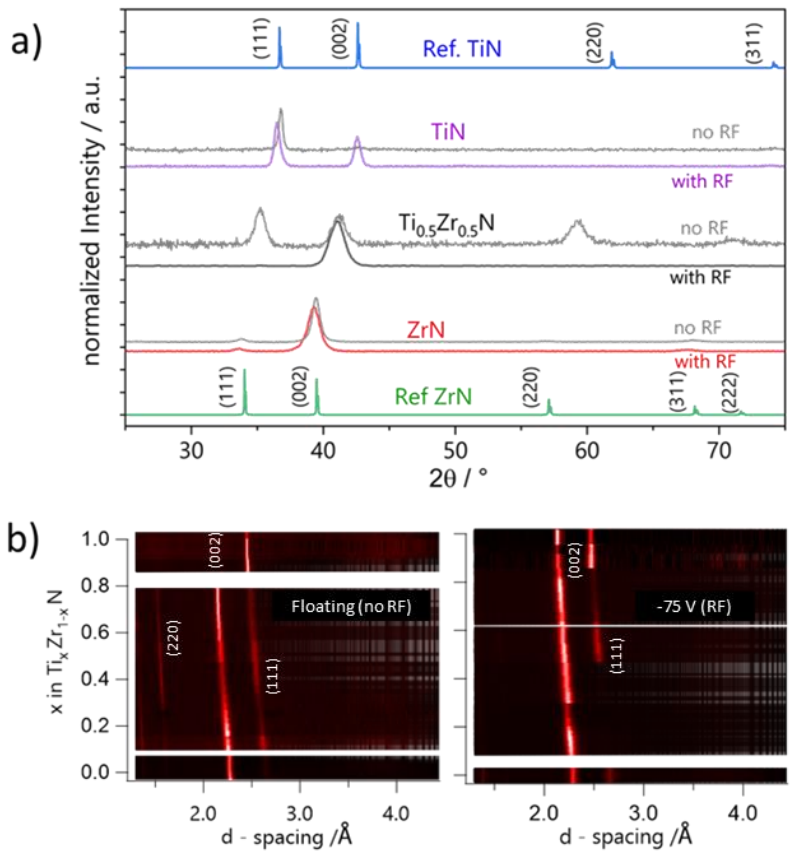

**Figure 2:** (a) Single XRD patterns of sputtered TiN, ZrN, and $Ti_{0.5}Zr_{0.5}N$ with reference diffraction patterns for TiN and ZrN. (b) *d*-spacing versus film thickness normalized intensity (background subtracted) of $Ti_xZr_{1-x}N$ coatings grown at floating potential and at -75 V RF substrate bias.

**Figure 2b** illustrates the XRD intensity as a function of cation composition and d-spacing normalized to layer thickness for the case of floating (Fig. 2 b left) and RF bias (Fig. 2 b right) substrate potentials. While the Zr atoms with a larger ionic radius (0.84 Å) are gradually incorporated into TiN, they replace the Ti atoms with a smaller ionic radius (0.61 Å). This explains the increase of the d spacing of the (002) from 2.121 Å (for TiN) to 2.294 Å (for ZrN) and for the (111) d-spacing the observed shift from 2.449 Å (TiN) to 2.649 Å (ZrN) (Fig. 5b). The change in Ti content leads to a gradual shift of the diffraction peaks towards higher angles, implying reduced d-spacing. Additionally, the diffraction peaks of $Ti_xZr_{1-x}N$ fall between those of TiN and ZrN, signifying that prepared $Ti_xZr_{1-x}N$ thin films crystallize in NaCl structure (SG225) like TiN and ZrN. The continuous d-spacing transition with film composition 0<x<1 in $Ti_xZr_{1-x}N$ in both scenarios is well-aligned with Vegardian lattice scaling indicating solid-solution alloying with no visible precipitation of secondary phases[36]. The observed peaks also suggest a predominantly (002)-oriented growth, as anticipated for these growth conditions. This inclination is more prominent in RF-biased deposition, while films synthesized without the RF substrate bias also display intensities for other reflections. In addition, to lack of energetic ion-irradiation during growth, it is likely that the higher oxygen content in the films deposited without the RF substrate bias also contributes to a reduced texture of the films.

## 3.2 Analysis of Functional Properties

### Electrical Properties

The observed changes in the structural properties and most importantly the difference in oxygen content with the application of the RF substrate bias are expected to have a pronounced effect on the functional properties of the material. First the resistivity was measured for combinatorial TiN and ZrN libraries featuring orthogonal gradients of deposition temperature and sputter flux. The libraries were deposited using a power density of 9.88 W/cm$^2$ at a pressure of 0.5 Pa in an Ar/N$_2$ gas flow ratio of 2/1. The libraries were deposited in an oblique angle setup with only one active magnetron resulting in the observed deposition rate gradient. These libraries represent a relatively broad variation of deposition conditions that could be employed in a standard research sputter chamber. **Figure 3** illustrates the resistivity as well as the level of oxygen contamination for combinatorial libraries of TiN and ZrN. The dots mark the 45 samples on the library. In this illustration, the color of the circles represents the resistivity, with darker circles indicating lower resistivity levels. The circle's size conveys the extent of oxygen contamination, as determined through semi-quantitative XPS analysis after sputter depth profile mapping. Larger circles signify a higher bulk oxygen content.

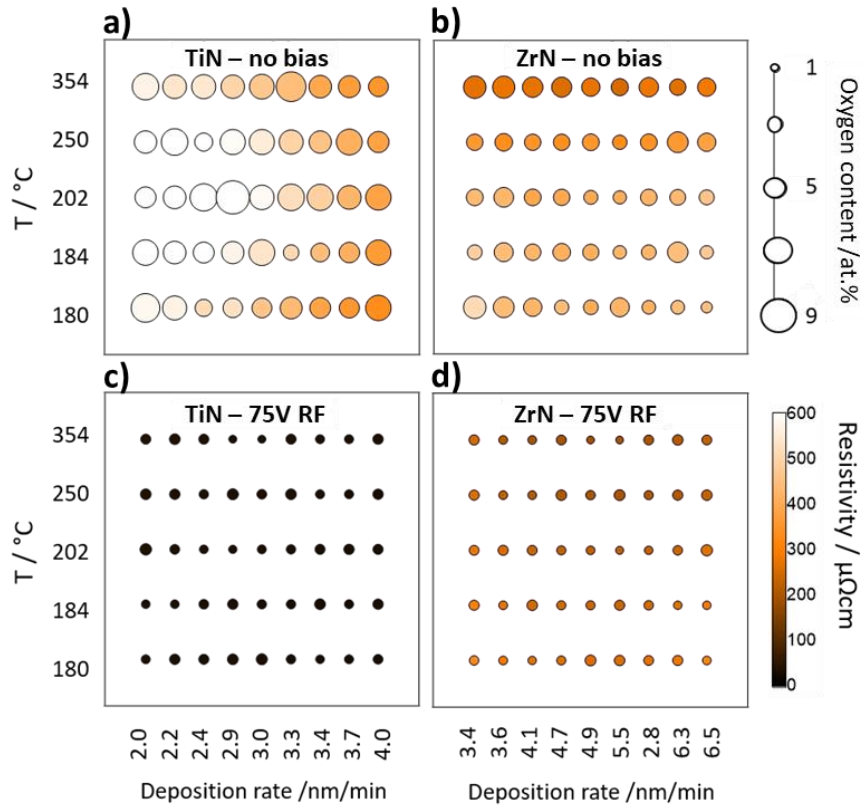

**Figure 3:** Libraries of deposited TiN (a,c) and ZrN (b,d) combinatorial libraries prepared without (a,b) or with an RF substrate bias (c,d). The y-axis shows a temperature gradient, the x-axis shows a variation in the deposition rate. The dots mark the 45 samples on the combinatorial library. The color scale shows the conductivity for each sample. The size of the circles reflects the oxygen contamination from XPS sputter profile mapping at that position.

For films grown without the RF substrate bias, there is an overall trend towards higher oxygen contamination in layers grown with lower deposition rate and lower substrate temperature, which is not surprising assuming a finite arrival rate of oxygen combined with high sticking coefficients and low desorption rates at these parameters. It has to be noted that the oxygen contamination in the chamber was not high and comparable to a standard UHV sputter tool. Neither the residual oxygen in the chamber was detectable by optical emission spectroscopy (OES) nor the negatively charged oxygen ions could be measured in a retarding field energy analyzer (RFEA). With increasing oxygen contamination in the films, resistivity increases, which is in qualitative agreement with previous studies. [37] The films deposited without RF substrate bias also show some variation of the resistivity with changes in deposition temperature and deposition rate. This can have multiple causes. TiN films with N content of less than 50 at% show increased resistivity linked to a random distribution of nitrogen over the interstitial sites, together with scattering at grain boundaries.[38] In addition, off-stoichiometry promotes the incorporation of impurities which can also act as recombination sites. Strikingly, the resistivity reduces significantly with the application of the substrate bias for both materials. To investigate whether this is mainly due to a reduction in oxygen contamination or changes in microstructure the correlation of crystal size and resistivity was also investigated. No discernible correlation between crystal size and resistivity was found in this study (see **Figure S5**). No clear correlation between preferred orientation as well as crystallinity and films' resistivity was also reported in previous studies.[30,31] This implies that the changes in resistivity are largely governed by the degree of oxygen contamination in the films.

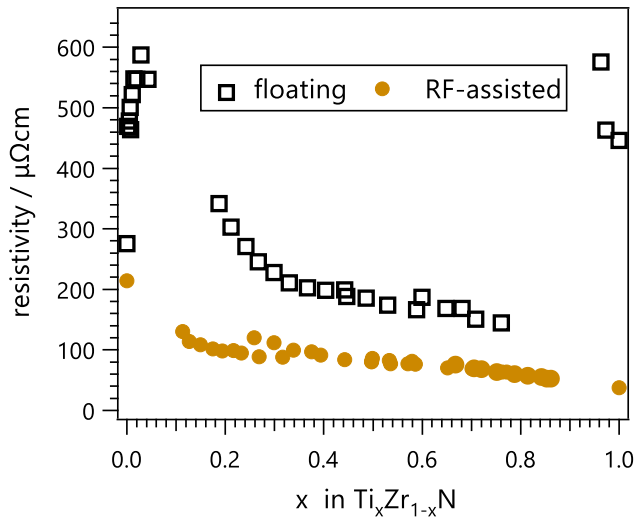

**Figure 4**: Resistivity of ternary $Ti_xZr_{1-x}N$ and binary films with (brown circles) and without (black squares) applied RF substrate bias.

The electrical properties of $Ti_xZr_{1-x}N$ thin films are also affected by the film's composition. **Figure 4** shows results from resistivity mapping on $Ti_xZr_{1-x}N$ combinatorial libraries for a substrate temperature of 354 °C. To minimize the effect of the deposition rate on the resistivity the nitrogen partial pressure was optimized for the functional property mapping on $Ti_xZr_{1-x}N$. With the optimized conditions (**see Table 1**) the measured resistivity variation was much smaller and nearly independent from the deposition rate (**Figure S4**) even without RF substrate bias. The partial pressure was held constant for all compositions, which resulted in higher resistivity for ZrN, which typically requires higher nitrogen partial pressure than TiN.

The resistivity of the films initially increases in films with x up to 0.05 and then decreases with increasing Ti content. Again, a significant effect of the RF substrate bias is visible, with the strongest difference for the pure binary compositions. The key difference between the two datasets is the level of oxygen contamination (see oxygen contamination studies section above).

The results in both Figure 3 and Figure 4 clearly show that the resistivity is largely governed by oxygen contamination in the films. TiN as well as ZrN deposited with an RF substrate bias not only show significantly reduced resistivity. The resistivity also becomes a lot less affected by changes in the process parameters, i.e. deposition temperature and deposition rate. Similarly the changes in resistivity remain small over large compositional regions when an RF substrate bias is applied, which facilitates compositional changes without compromising on the electrical properties. This highlights the potential of the technique for developing not only highly performant coatings, but also robust and reproducible deposition processes.

## Mechanical properties

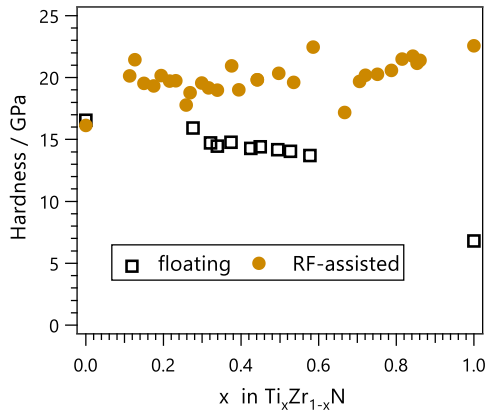

**Figure 5:** Nanoindentation hardness of the ternary $Ti_xZr_{1-x}N$ coatings as a function of x. Values for binaries are provided for the films grown with deposition rates of 3.4 nm/min, close to the average deposition rate of the combinatorial libraries.

To investigate the influence of the RF substrate bias on the mechanical properties of the films nano-indentation was performed on TiN, ZrN as well as $Ti_xZr_{1-x}N$ libraries. **Figure 5** shows the nano-hardness for different compositions and a deposition temperature of 354 °C. Only samples with a constant deposition rate were selected to increase the reproducibility of the study. A clear effect of the RF substrate bias during the deposition of $Ti_xZr_{1-x}N$ films on their mechanical properties is visible in Figure 5. The RF-biased samples exhibit a higher hardness of around 20 GPa throughout the entire compositional range, whereas unbiased samples exhibit a clear decline towards high Ti content. Zr has a larger atomic radius (0.160 nm) than Ti (0.147 nm), which may affect distortion when incorporated in TiN. Strikingly, no solid solution hardening is observed with incorporation of Zr into the TiN lattice.[39,40] However, the deposition process was optimized for TiN. ZrN requires a higher N partial pressure. Thus ZrN films are sub-stoichiometric according to XPS, where typical measured values for Zr:N were 55:45 in contrast to Ti:N = 48:52; for ternary samples, this meant that as soon as Zr content was increased, N content lowered and the films became sub-stoichiometric. A higher $N_2$ partial pressure will however result in over-stoichiometric TiN films with a brownish color, indicating a softer TiN coating [41]. Without RF bias, for TiN and Ti-rich $Ti_xZr_{1-x}N$ the hardness was found to drop significantly. This can be attributed to the increased oxygen content and the potential formation of oxynitrides and oxide secondary phases. Typically, for oxynitrides, such as TiAlNO, VAlNO, and VMoNO, due to charge balancing, the metallic bonding (soft) character increases at the expense of stronger covalent bonds with increasing oxygen content.[42–44] This makes oxynitrides softer than nitrides. Also, Me-O bonds are longer than Me-N, inferring a weaker bond. Hence, with increasing oxygen content the hardness drops. These changes in hardness can also be attributed to the formation of nitrogen vacancies in the TiZrN lattice, which can act as sources of plastic deformation and lead to a decrease in the hardness and Young's modulus.[45] While the slight variation in cation/anion ratio complicated a clear assessment of the hardness as a function of the composition, the effect of the RF substrate bias is remarkable.

In addition to a change in hardness the films, a change in the microstructure of the growing film (with and without applied RF bias) is also indicated by the change of roughness of the films. TiN films deposited with RF substrate bias were measured at 1.4 nm RMS, and without bias at 5.7 nm Rms, as depicted in Fig.S6. This change in surface roughness can be attributed to changes in grain size. With the applied RF substrate bias, the adatom mobility is higher due to increased energy input from the plasma via several processes.[46] In addition, the oxygen contamination is lower minimizing local lattice distortion. Both effects contribute to the growth of compact films with high crystalline quality. Consequently, it appears that the application of the RF bias has several benefits positively influencing phase-purity, but also microstructure and crystallinity.

## 4. Conclusion

The incorporation of light element contaminants is a key challenge in the sputter deposition of thin films. Oxygen contamination is known to adversely affect the properties of many functional thin film materials, in particular nitrides. Many potential avenues exist to improve film purity, but they are typically associated with higher investment and production costs. In this work we show how RF substrate bias potentials can be used to significantly reduce, if not eliminate oxygen contamination in reactively sputtered nitride thin films. We use TiN and ZrN as prototypical systems to demonstrate the effectiveness of the method. In addition, a combinatorial approach is used to demonstrate the feasibility for a wide variety of compositions in the (Ti,Zr)N phase system.

We show that a low bias voltage value can be exploited to control the oxygen contamination. Ideally, the substrate bias voltage should be chosen so that the ion energy is higher than the nitrogen sputtering threshold and lower than the argon sputtering threshold. In this way, the adsorbed oxygen is preferentially sputtered. In all investigated materials applying a slight RF bias to the substrate is shown to dramatically reduce the oxygen incorporation.

This in turn, leads to significantly improved structural and functional properties of the materials. XRD patterns showed that the crystallinity of the films improved with the application of a RF substrate bias. AFM images showed that the surface morphology of the films improved, and the surface roughness decreased with the application of a RF substrate bias. Four-point probe measurements showed that the electrical resistivity of the films decreased with the application of a RF substrate bias, while nanoindentation measurements showed an increase in hardness. These results demonstrate that the application of a RF bias to the substrate during DC sputtering can be an effective approach to improve the properties of thin films deposited on insulating substrates.

The approach is applicable to insulating substrates and conductive substrates. In addition, it is likely transferrable to many different materials and holds the potential to reduce production costs for functional nitride coatings in the future as well as increase the reproducibility of the experimental results in varying process environments.

## Acknowledgements

The authors would like to acknowledge help by Alexander Wieczorek with the data analysis in CombIgor, Monalisa Ghosh for help with the OES measurements as well as Ulrich Müller for support with the deposition experiments. Giacomo Lorenzin and Claudia Cancellieri are gratefully acknowledged for providing the additional TiN films deposited in the AJA Orion 8 tool. Funding by the SNF (Grant No. 196980) as well as Innosuisse (Grant No. 50764.1 IP-EE) are gratefully acknowledged. S.Z. acknowledges research funding from Empa Internal Research Call 2020. Alexander Groetsch is gratefully acknowledged for his contributions to the development of the mapping nanoindenter.

## Author Contributions

**K.T.:** Conceptualization, Investigation, Methodology, Formal analysis, Visualization, Writing – Original Draft; **M.W.:** Investigation, Formal analysis, Writing – Review & Editing; **O.P.:** Investigation, Formal analysis, Visualization, Writing – Review & Editing; **S.Z.:** Investigation, Writing – Review & Editing; **J.P.:** Investigation, Writing – Review & Editing; **J.Sc.:** Supervision, Writing – Review & Editing; **J.So.:** Investigation, Writing – Review & Editing; **L.S.:** Funding acquisition, Writing – Review & Editing; **S.S.:** Conceptualization, Supervision, Methodology, Funding acquisition, Writing – Review & Editing

## Conflict of Interest

The authors declare no conflict of interest.

# References

1. Alberi, K. *et al.* The 2019 materials by design roadmap. *J. Phys. D. Appl. Phys.* **52**, 013001 (2019).

2. Sun, W. *et al.* A map of the inorganic ternary metal nitrides. *Nat. Mater.* **18**, 732–739 (2019).

3. Green, M. L. *et al.* Fulfilling the promise of the materials genome initiative with high-throughput experimental methodologies. *Appl. Phys. Rev.* **4**, 011105 (2017).

4. Greene, J. E. Review Article: Tracing the recorded history of thin-film sputter deposition: From the 1800s to 2017. *J. Vac. Sci. Technol. A Vacuum, Surfaces, Film.* **35**, 0–60 (2017).

5. Hultman, L. Thermal stability of nitride thin films. *Vacuum* **57**, 1–30 (2000).

6. You, A., Be, M. A. Y. & In, I. On structure and properties of sputtered Ti and Al based hard compound films. **2695**, (2003).

7. Prange, R., Cremer, R. U. & Neuschutz, D. Plasma-enhanced CVD of ž Ti , Al / N films from chloridic precursors in a DC glow discharge. 208–214 (2000).

8. Cremer, R. & Neuschutz, D. A combinatorial approach to the optimization of metastable multicomponent hard coatings. **147**, 229–236 (2001).

9. Group, T. F. & Technology, M. Review Paper. **128**, 21–44 (1985).

10. Wittmer, M., Noser, J. & Melchior, H. I-. **6664**, 6659–6664 (1981).

11. Akhadejdamrong, T. *et al. Formation of Protection Layer during Oxidation of Al-Implanted TiN Coating*. *Materials Transactions* vol. 43 (2002).

12. Hollerweger, R. *et al.* Guidelines for increasing the oxidation resistance of Ti-Al-N based coatings. *Thin Solid Films* **688**, 137290 (2019).

13. Diserens, M., Patscheider, J. & Lévy, F. Improving the properties of titanium nitride by incorporation of silicon. *Surf. Coatings Technol.* **108**–**109**, 241–246 (1998).

14. Vennemann, A., Stock, H. R., Kohlscheen, J., Rambadt, S. & Erkens, G. Oxidation resistance of titanium–aluminium–silicon nitride coatings. *Surf. Coatings Technol.* **174**–**175**, 408–415 (2003).

15. Takeyama, M. B., Itoi, T., Aoyagi, E. & Noya, A. Diffusion barrier properties of nano-crystalline TiZrN films in Cu/Si contact systems. in *Applied Surface Science* vol. 216 181–186 (Elsevier, 2003).

16. Jeon, S., Ha, J., Choi, Y., Jo, I. & Lee, H. Interfacial stability and diffusion barrier ability of Ti 1-x Zr x N coatings by pulsed laser thermal shock. *Appl. Surf. Sci.* **320**, 602–608 (2014).

17. Ningthoujam, R. S. & Gajbhiye, N. S. Synthesis, electron transport properties of transition metal nitrides and applications. *Prog. Mater. Sci.* **70**, 50–154 (2015).

18. Nagakura, S., Hirotsu, Y. & Vt, V. J. J. Lattice parameter of the non-stoichiometric compound $TiN_x$. *J. Appl. Cryst.* 156–157 (1972).

19. De Maayer, P. J. P. & Mackenzie, J. D. The Electrical Properties of Thin Films of $TiN_x$ and $TiC_x$. *Zeitschrift fur Naturforsch. - Sect. A J. Phys. Sci.* **30**, 1661–1666 (1975).

20. Signore, M. A. *et al.* Investigation of the physical properties of ion assisted ZrN thin films deposited by RF magnetron sputtering. *J. Phys. D. Appl. Phys.* **43**, (2010).

21. Jeyachandran, Y. L., Narayandass, S. K., Mangalaraj, D., Areva, S. & Mielczarski, J. A. Properties of titanium nitride films prepared by direct current magnetron sputtering. *Mater. Sci. Eng. A* **445–446**, 223–236 (2007).

22. Martin, N. *et al.* Correlation between processing and properties of TiO x N y thin films sputter deposited by the reactive gas pulsing technique. *Appl. Surf. Sci.* **185**, 123–133 (2001).

23. Westra, K. L., Lawson, R. P. W. & Brett, M. J. The effects of oxygen contamination on the properties of reactively sputtered indium nitride films. *J. Vac. Sci. Technol. A Vacuum, Surfaces, Film.* **6**, 1730–1732 (1988).

24. Maissel, L. I. & Schaible, P. M. Thin films deposited by bias sputtering. *J. Appl. Phys.* **36**, 237–242 (1965).

25. Zhuk, S. *et al.* Combinatorial Reactive Sputtering with Auger Parameter Analysis Enables Synthesis of Wurtzite Zn2TaN3. *Chem. Mater.* (2023) doi:10.1021/acs.chemmater.3c01341.

26. Trant, M. *et al.* Tunable ion flux density and its impact on AlN thin films deposited in a confocal DC magnetron sputtering system. *Surf. Coatings Technol.* **348**, 159–167 (2018).

27. Anders, A. A structure zone diagram including plasma-based deposition and ion etching. *Thin Solid Films* **518**, 4087–4090 (2010).

28. Oliver, W. C. & Pharr, G. M. Measurement of hardness and elastic modulus by instrumented indentation: Advances in understanding and refinements to methodology. *J. Mater. Res.* **19**, 3–20 (2004).

29. Nah, J.-W., Kim, B.-J., Lee, D.-K. & Lee, J.-J. Color, structure, and properties of TiN coatings prepared by plasma enhanced chemical vapor deposition. *J. Vac. Sci. Technol. A Vacuum, Surfaces, Film.* **17**, 463–469 (1999).

30. Nose, M. *et al.* Electrical and colorimetric properties of TiN thin films prepared by DC reactive sputtering in a facing targets sputtering (FTS) system. *Surf. Coatings Technol.* **116–119**, 296–301 (1999).

31. Nose, M., Zhou, M., Honbo, E., Yokota, M. & Saji, S. Colorimetric properties of ZrN and TiN coatings prepared by DC reactive sputtering. *Surf. Coatings Technol.* **142–144**, 211–217 (2001).

32. Pshyk, A. V. *et al.* Energy-efficient physical vapor deposition of dense and hard Ti-Al-W-N coatings deposited under industrial conditions. *Mater. Des.* **227**, (2023).

33. Patidar, J. *et al.* Improving the crystallinity and texture of oblique-angle-deposited AlN thin films using reactive synchronized HiPIMS. *Surf. Coatings Technol.* **468**, 129719 (2023).

34. Jain, A. *et al.* Commentary: The Materials Project: A materials genome approach to accelerating materials innovation. *APL Mater.* **1**, 011002 (2013).

35. Abadias, G., Tse, Y. Y., Guérin, P. & Pelosin, V. Interdependence between stress, preferred orientation, and surface morphology of nanocrystalline TiN thin films deposited by dual ion beam sputtering. *J. Appl. Phys.* **99**, (2006).

36. King, H. W. & Vegard, Y. Lo. Yon Lo Vegard. *J. Mater. Sci.* **1**, 79–90 (1921).

37. Nakano, T., Hoshi, K. & Baba, S. Effect of background gas environment on oxygen incorporation in TiN films deposited using UHV reactive magnetron sputtering. *Vacuum* **83**, 467–469 (2008).

38. Kurtz, S. R. & Gordon, R. G. Chemical vapor deposition of titanium nitride at low temperatures. *Thin Solid Films* **140**, 277–290 (1986).

39. Herrera-Jimenez, E. J. *et al.* Solid solution hardening in nanolaminate ZrN-TiN coatings with enhanced wear resistance. *Thin Solid Films* **688**, 137431 (2019).

40. Uglov, V. V., Anishchik, V. M., Zlotski, S. V., Abadias, G. & Dub, S. N. Structural and mechanical stability upon annealing of arc-deposited Ti–Zr–N coatings. *Surf. Coatings Technol.* **202**, 2394–2398 (2008).

41. Sproul, W. D., Rudnik, P. J. & Gogol, C. A. *THE EFFECT OF TARGET POWER ON THE NITROGEN PARTIAL PRESSURE LEVEL AND HARDNESS OF REACTIVELY SPUTTERED TITANIUM NITRIDE COATINGS*. *Thin Solid Films* vol. 171 (1989).

42. Baben, M. *et al.* Phase stability and elastic properties of titanium aluminum oxynitride studied by ab initio calculations. doi:10.1088/0022-3727/46/8/084002.

43. Hans, M. *et al.* Effect of oxygen incorporation on the structure and elasticity of Ti-Al-O-N coatings synthesized by cathodic arc and high power pulsed magnetron sputtering Effect of oxygen incorporation on the structure and elasticity of Ti-Al-O-N coatings synthesized by. **093515**, (2014).

44. Edström, D. *et al.* development of hard and tough refractory oxynitrides Mechanical properties of VMoNO as a function of oxygen concentration : Toward development of hard and tough refractory oxynitrides. **061508**, (2019).

45. Tsai, D. C. *et al.* Wide variation in the structure and physical properties of reactively sputtered (TiZrHf)N coatings under different working pressures. *J. Alloys Compd.* **750**, 350–359 (2018).

46. Anders, A. A structure zone diagram including plasma-based deposition and ion etching. *Thin Solid Films* **518**, 4087–4090 (2010).

# Reducing the oxygen contamination in conductive (Ti,Zr)N coatings *via* RF-bias assisted reactive sputtering

*K. Thorwarth[1]\*, M. Watroba[2], O. Pshyk[1], S. Zhuk[1], J. Patidar[1], J. Schwiedrzik[2], J. Sommerhäuser[1], L. Sommerhäuser[1], S. Siol[1]\**

[1] Empa, Swiss Federal Laboratories for Materials Science and Technology, Laboratory for Surface Science and Coating Technologies, Dübendorf, Switzerland

[2] Empa, Swiss Federal Laboratories for Materials Science and Technology, Laboratory for Mechanics of Materials and Nanostructures, Thun, Switzerland

*E-Mail: kerstin.thorwarth@empa.ch, sebastian.siol@empa.ch*

KEYWORDS TiN, ZrN, Oxygen Contamination, Magnetron Sputtering, Combinatorial Screening, Substrate Bias Potential

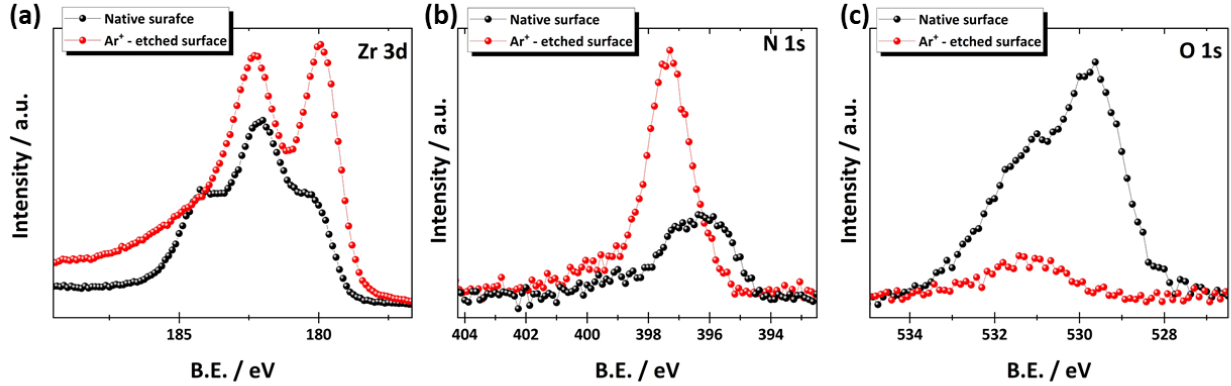
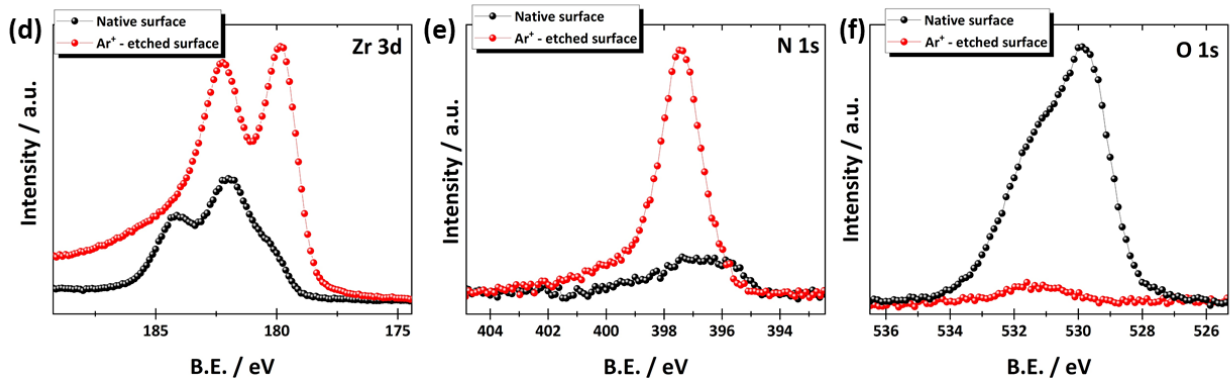

**Fig. S1:** XPS core-level Zr 3d, N 1s, and O 1s spectra acquired before (black lines and symbols) and after (red lines and symbols) Ar+ sputter-etching of ZrN films deposited without (a-c) and with (d-f) RF substrate bias.

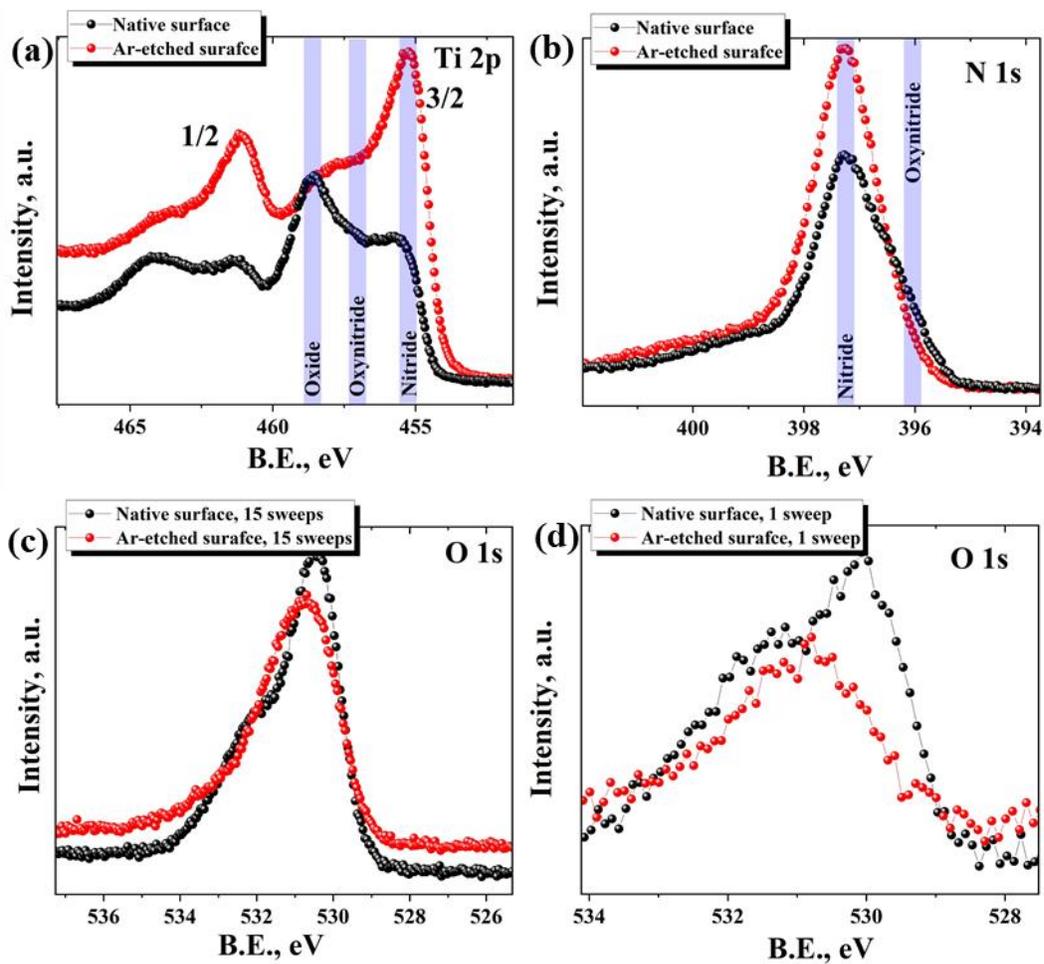

**Fig. S2:** Ti 2p, N 1s, and O 1s XPS core-level spectra acquired before (black lines and symbols) and after (red lines and symbols) Ar⁺ sputter-etching of the reference TiN films deposited without RF substrate bias in an additional AJA Orion deposition system.

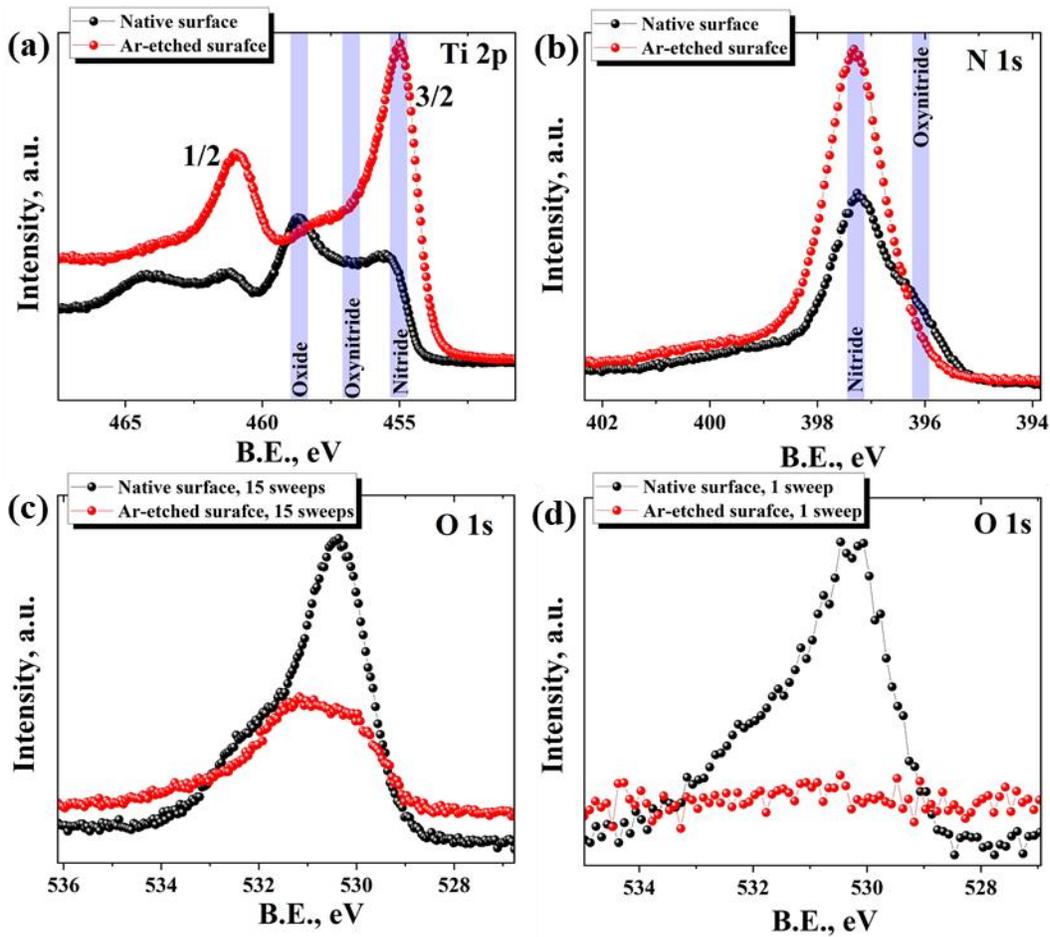

**Fig. S3:** Ti 2p, N 1s, and O 1s XPS core-level spectra acquired before (black lines and symbols) and after (red lines and symbols) Ar$^+$ sputter-etching of the reference TiN films deposited with RF substrate bias in an additional AJA Orion deposition system.

**Table S1:** XPS-derived composition of TiN films grown in an additional AJA Orion deposition system without and with RF substrate bias.

| Sample | | Ti, at % | N, at % | O, at % | C, at % |
| --- | --- | --- | --- | --- | --- |
| TiN-without RF bias, ambient transfer, Orion 8 | Native surface | 23.7 | 25.0 | 28.3 | 23.0 |
| | Ar$^+$ sputter-etched | 42.1 | 39.7 | 13.7 | 4.5 |
| TiN-RF bias, ambient transfer, Orion 8 | Native surface | 25.0 | 24.5 | 25.3 | 25.2 |
| | Ar$^+$ sputter-etched | 50.0 | 48.6 | 0.4 | 1.0 |

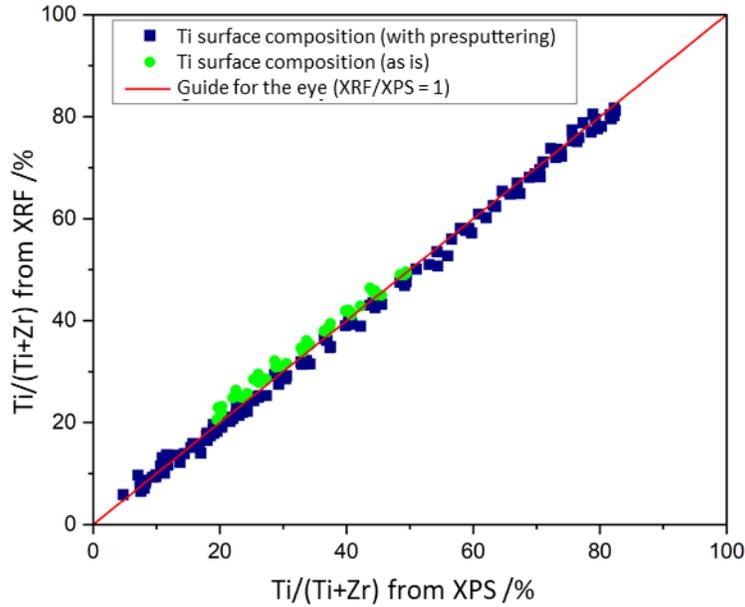

**Figure S4:** Comparison of Ti content determined by XRF with the Ti content measured by XPS. The surface concentration of Ti slightly higher but within error.

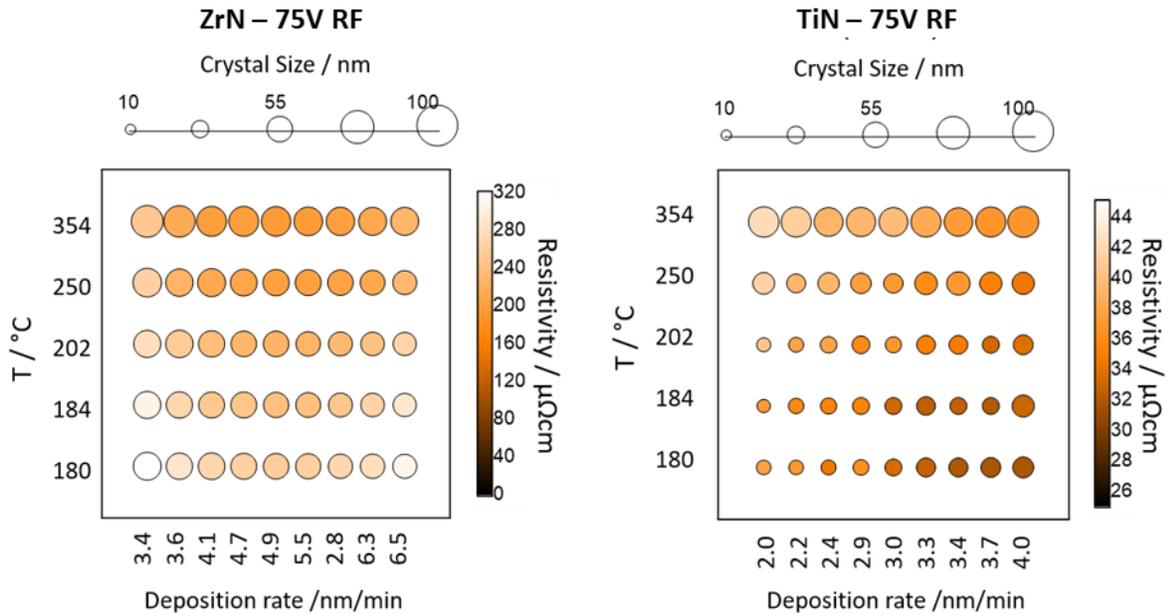

**Figure S5:** resistivity (indicated by the color) of a ZrN (left) and a TiN (right) library. On the c-axis, crystallite size is indicated by the size of the circles. The x-axis indicates a change in the deposition rate, the y-axis a temperature gradient. No pronounced correlation between crystal size and resistivity is observed.

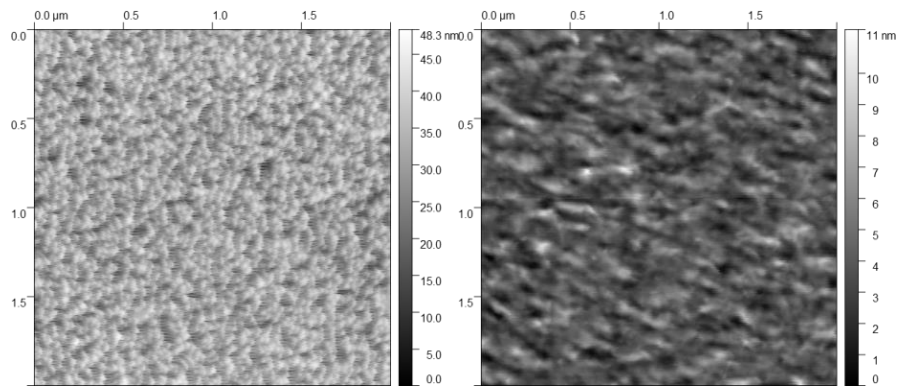

**Figure S6:** Layer morphologies for TiN coatings deposited without (left) and with (right) RF bias.

The surface roughness and morphology of TiZrN thin films were observed using an atomic force microscope (AFM) (Icon 3, Bruker) in ScanAsyst mode with Silicon cantilever tips. The 2 × 2 µm² AFM images with 512 × 512-pixel resolution were post-processed in Gwyddion software.